\documentclass[aip,rsi,reprint,graphicx]{revtex4-1} 
\usepackage{graphicx} 
\usepackage{booktabs} 
\usepackage{multirow}
\usepackage{amsmath}
\usepackage{pbox}
\usepackage[ngerman, english]{babel}
\usepackage{subcaption}

\draft 

\begin{document}


\title{Magnetic Flux Tailoring through Lenz Lenses in Toroidal Diamond Indenter Cells: A New Pathway to High Pressure Nuclear Magnetic Resonance }  



\author{Thomas Meier}
\email[]{thomas.meier@uni-bayreuth.de}
 \affiliation{Bayerisches Geoinstitut, Bayreuth University, Universit{\"a}tsstra\ss e 30, 95447 Bayreuth, Germany}
 
\author{Nan Wang}
\affiliation{Institute of Microstructure Technology, Karlsruhe Institute of Technology, Hermann-von-Helmholtz-Platz 1, 76344 Eggenstein-Leopoldshafen, Germany}

\author{Dario Mager}
\affiliation{Institute of Microstructure Technology, Karlsruhe Institute of Technology, Hermann-von-Helmholtz-Platz 1, 76344 Eggenstein-Leopoldshafen, Germany}

\author{Jan G. Korvink}
\affiliation{Institute of Microstructure Technology, Karlsruhe Institute of Technology, Hermann-von-Helmholtz-Platz 1, 76344 Eggenstein-Leopoldshafen, Germany}

\author{Sylvain Petigirard}
\affiliation{Bayerisches Geoinstitut, Bayreuth University, Universit{\"a}tsstra\ss e 30, 95447 Bayreuth, Germany}

 \author{Leonid Dubrovinsky}
  \affiliation{Bayerisches Geoinstitut, Bayreuth University, Universit{\"a}tsstra\ss e 30, 95447 Bayreuth, Germany}


\date{\today}

\begin{abstract}
A new pathway to nuclear magnetic resonance spectroscopy in high pressure diamond anvil cells is introduced, using inductively coupled broadband passive electro-magnetic lenses to locally amplify the magnetic flux at the isolated sample, leading to an increase in sensitivity. The lenses are adopted for the geometrical restrictions imposed by a toroidal diamond indenter cell, and yield high signal-to-noise ratios at pressures as high as 72 GPa, at initial sample volumes of only 230 pl. The corresponding levels of detection, $\text{LOD}_t$, are found to be up to four orders of magnitude lower compared to formerly used solenoidal micro-coils in diamond anvil cells, as shown by $^1$H-NMR measurements on paraffin oil.
This approach opens up the field of ultra-high pressure sciences for one of the most versatile spectroscopic methods available in a pressure range unprecedended up to now.
\end{abstract}

\maketitle

\section{Introduction}

Nuclear magnetic resonance (NMR) spectroscopy is by far the most widespread analytical method in modern life science. Especially biology, chemistry, and also medicine are benefiting from NMR's ability to locally yield valuable structural, electronic, and dynamical information; and it is used by an ever growing community spanning almost all the of natural sciences\cite{Grant2007, Ernst2010, Andrew1984}. Especially since the development of \textit{in-vivo} magnetic resonance imaging\cite{Lauterbur1973, Mansfield2004} and due to its singular analytical role in the investigation of proteins\cite{Riek1996, Zahn2000, Scheidt2011}, NMR has become an integral part in these research fields.\\
 \begin{figure}[htbp]
 \centering
 \includegraphics[scale=0.18]{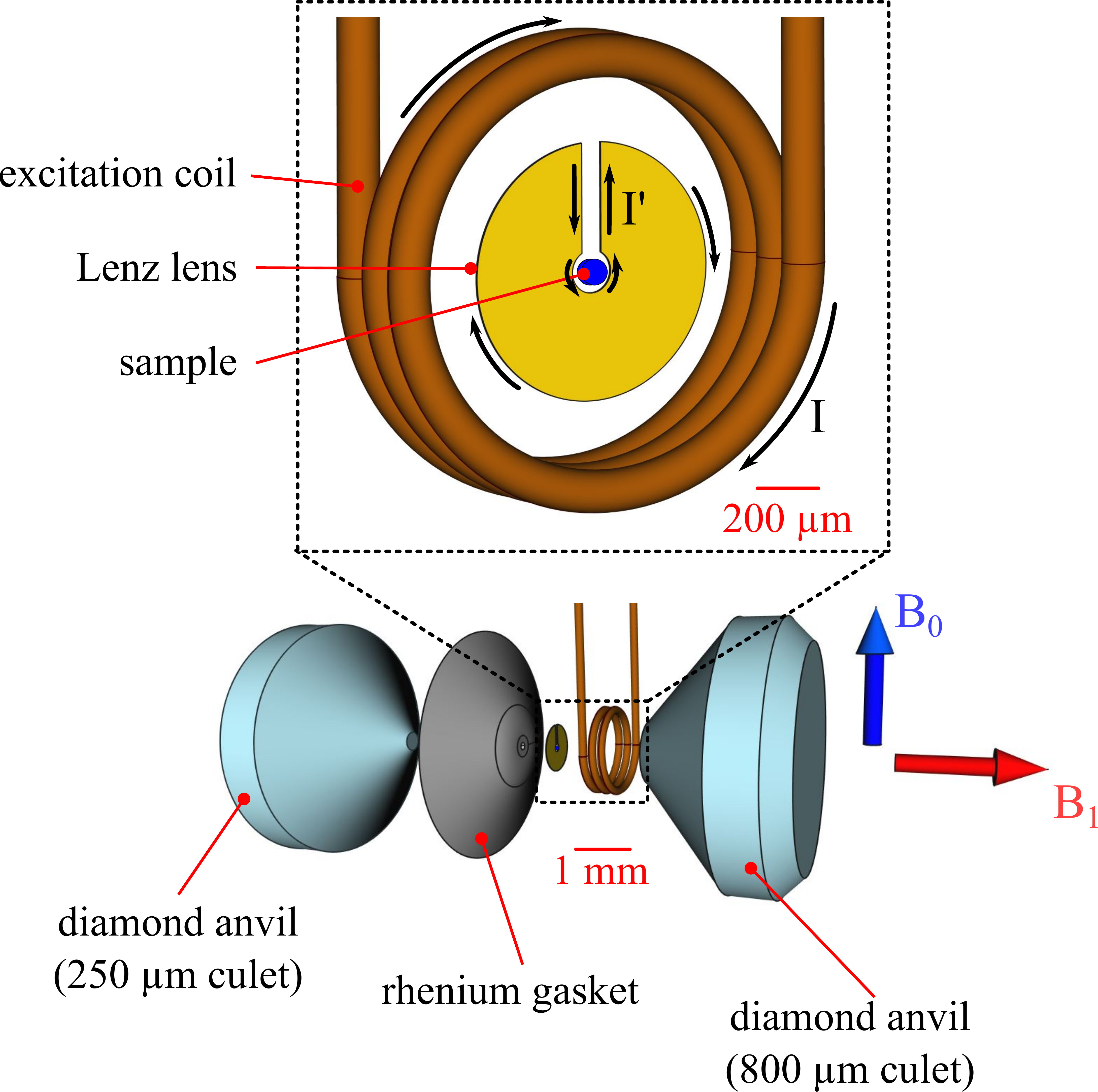}%
 \caption{Schematic explosion diagramme of the resonator set-up and the anvil/gasket arrangement.The blue and red arrows denote the directions of the external magnetic field B$_0$ and the r.f. magnetic field B$_1$ generated by the excitation coil and the lens. The enlarged picture shows the r.f. arrangement of the excitation coil with the Lenz lens. Black arrows denote the directions of the high-frequency current in the excitation coil, $I$, as well as the induced current in the lens, $I'$.\label{fig1}}
 \end{figure}
Besides its widespread use, there are some research branches where a broader application of NMR seems unfeasible, for example in experimental geosciences or high pressure chemistry, where harsh experimental conditions such as high pressure and high temperature are mandatory. Nevertheless, NMR under such harsh conditions will certainly have a great impact on modern high pressure chemistry and geosciences, where an application of NMR is mostly restricted to \textit{ex-situ} measurements on recovered samples\cite{Yarger1995a, Stebbins1995}. Additionally, it is believed that first signs of life on Earth developed in the considerable depths of the proto-ocean of the hadean era, under high pressure conditions\cite{Sharma2002, Daniel2006, Huang2016}. Moreover, it was realised early on that pressure, as a thermodynamic parameter, can be used to elucidate protein structure and function at ambient conditions\cite{Bridgman1914, Balny2006, Heremans1998, Silva2001, Mozhaev1996, Royer2002, Golub2017}.\\
The main obstacle against an application of high pressure NMR lie in its inherently low sensitivity and the requirement that the radio frequency (r.f.) transceiver must be very close to the sample, ensuring a good filling factor of the resonator\cite{Fukushima1994}. In high pressure generating vessels, i.e. a diamond anvil cells, typical available sample spaces are less than 1 nl, and is also tightly enclosed by two diamond anvils and a metallic gasket; thus, the use of \textit{in-situ} NMR experiments was widely considered an impossibility for several decades. \\
Nonetheless, several research groups were able to implement NMR in diamond anvil cells at pressures up to 10 GPa (1 GPa = 10.000 bar)\cite{Meier2017}. These previous set-ups suffered from low sensitivites and therefore were only applicable to systems rich in \grqq high-$\gamma_n$\grqq nuclei, such as hydrogen or fluorine. A turning point for high pressure NMR research was the application of resonating multi-turn micro-coils. In 2009, Suzuki et al. introduced the idea to place such a minuscle micro-coil \---typically 400\--500 $\mu m$ in diameter and 100\--250 $\mu m$ in height\--- directly in the sample chamber between the anvil vise\cite{Suzuki2009}, and it could be shown that this technique was indeed superior to the previous attempts in terms of spin sensitivities\cite{Meier2014, Meier2015} and reachable pressures of up to 30 GPa\cite{Meier2015a}\\
Of course that bold approach came with some drawbacks. The extremely fragile micro-coils were made from very thin insulated gold or copper wire, and become exceedingly difficult to manufacture and to handle if a further miniaturisation is needed in order to reach pressures above 10 GPa\cite{Meier2014a} with an imposed empirical pressure limit of about 15 \-- 18 GPa.\\
Unfortunately, the most pressing scientific questions in high pressure chemistry and the geosciences nowadays concern topics such as high pressure phase transitions towards new exotic materials\cite{Schwarz1999, Bykova2016}, coordination changes or spin transitions of Earth's mantle materials\cite{Frost2004, Kupenko2015a, Kurnosov2017}, or metallisations or even transitions into a superconducting state of diatomic molecules such as hydrogen\cite{Eremets2011, Eremets2016, Dalladay-Simpson2016, Dias2017}; these all occur predominantly at much higher pressures close to the mega-bar regime (1 Mbar = 100 GPa) or even beyond. Therefore, in order to implement \textit{in-situ} NMR measurements at such extreme pressures, new radio frequency resonators must be developed, enabling a successful detection of the NMR signal from within the pressure chamber, yielding high sensitivities throughout the whole experiment.\\
\section{Results}
A possible solution was introduced using magnetic flux tailoring Lenz lenses, which were recognised recently to locally amplify the magnetic flux at a given region of interest by Schoenmaker et al.\cite{Schoenmaker2013}, and in more detail by Spengler et al.\cite{Spengler2016}.  Such lenses are typically made from thin wire or from a solid sheet of copper or gold, and their working principle is directly governed by Lenz' law of induction. It could be shown that these resonators are capable of focusing the total magnetic field of the resonator at its centre, leading to a locally enhanced sensitivity.\\
This article will demonstrate that inductively coupled Lenz lenses can be used in a 
toroidal diamond indenter cell (TDIC), figure \ref{fig1}. $^1$H-NMR spectra obtained of paraffin  up to 72 GPa show the lenses' applicability in high pressure NMR research. Furthermore, using numerical finite element simulations, it is be revealed that application of a quasi two-dimensional resonator is preferable to the bigger and more fragile solenoidal coils which have been used before.\\
Preparation of the TDIC pressure cell as well as the r.f. resonator are described in the online methods section.\\
In order to compare sensitivities and performances under pressure for different resonator types, it is instructive to define the limit of detection in the time domain, LOD$_t$, as the minimal necessary number of spins resonating within a bandwidth of 1 Hz yielding a signal-to-noise ratio of 1\cite{Ryan2012}:
\begin{equation}
\text{LOD}_t=\frac{\text{N}_{spins}}{\text{SNR}_t\cdot \sqrt{\Delta \text{f}}}
\label{LOD}
\end{equation}
where N$_{spins}$ denotes the number of spins contributing to the signal, SNR$_t$ the signal-to-noise ratio acquired in time domain and $\Delta$f the receiver bandwidth. Assuming that the sample chamber is filled completely with the liquid paraffin oil, the number of resonant spins can be gauged to be about $1.7\cdot 10^{16}$. 
 \begin{figure*}
  \includegraphics[scale=.25]{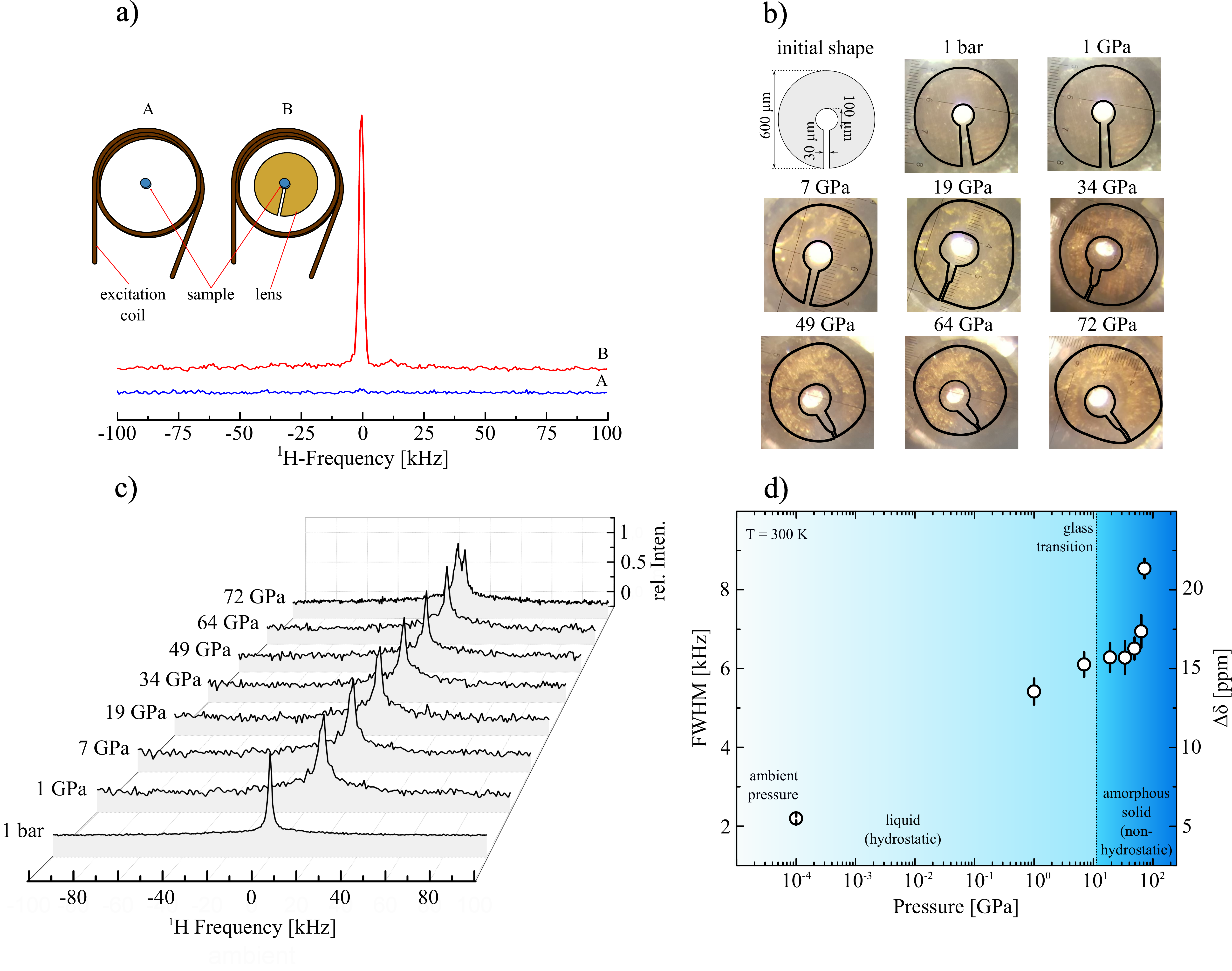}%
 \caption{a) Proton spectra of parrafin at ambient pressure with and without the use of a Lenz lens in a TDIC. b) Photographs of different deformation states of the lens under pressure. c) recorded $^1$H NMR spectra; at ambient conditions, 100 scans were accumulated whereas at higher pressures only single shot spectra after a single $\pi /2$-pulse were recorded. d) pressure dependence of the FWHM line widths; the dotted line denotes the crystallization pressure at ambient temperature, the shades areas denote the liquid as well as amorphous phases of paraffin. The glass transition pressure was obtained from other methods.\label{fig567}}%
 \end{figure*}
The effect of the Lenz lens on the sensitivity at ambient pressure in the TDIC is shown in figure \ref{fig567}a. The use of the lens strongly enhances the SNR and thus LOD$_t$ by about three orders of magnitude compared to the same arrangement measured without a lens. The bad performance without the lens partly originates in very poor filling factors of the outer excitation coil ($\eta\approx3\cdot10^{-4}$) and in B$_1$ field inhomogeneities at the sample. In fact, the \grqq lense-less\grqq~ arrangement is similar to previous attempts to perform high pressure NMR in DACs, yielding comparable sensitivities\cite{Pravica1998, Okuchi2005}.\\
The deformation of the lens under axial pressure is shown in figure \ref{fig567}b). Its overall shape remains stable at pressures up to about 7 GPa, after which the gold begins to deform up to about 20 GPa. A short cut, which would occur if the 30 $\mu m$ slit is closed, did not occur, possibly because some amount of the insulating Al$_2$O$_3$  layer between rhenium gasket and  Lenz lens (see online methods) had been squeezed into the slit, preventing a complete closure of the lens structure.
 The inner hole diameter was found to be increased by about 5\% at 72 GPa compared to ambient pressure together with a sliding of the hole away from the pressure centre by about 15 $\mu m$ originating from small misalignments of the anvil tips. Interestingly, the average r.f. field strength $\langle B_1\rangle$ produced by the lenses was found to be almost constant, as indicated by the obtained values from 2D nutation experiments, see table 1.  In principle, B$_1$ should scale with the inner diameter of the lens. Nevertheless, such an effect is most likely masked by the prescence of the metallic rhenium gasket which is in close proximity to the lenses and certainly also contribute somewhat to the total $\langle B_1\rangle$ field strength and homogeneity in the sample cavity.\\
\begin{table}
\begin{tabular}{ccccc}
p&t$_{\pi/2}$&$\langle B_1\rangle$&SNR$_t$&LOD$_t$\\
$[GPa]$&$[\mu s]$&$[mT]$& &[spins/$\sqrt{Hz}$] \\
\hline\\
\vspace{.3em}
10$^{-4}$&2.4&2.5&19&6$\cdot10^{11}$ \\ 
\vspace{.3em}
1&2.1&3.0&18.2&7$\cdot10^{11}$ \\ 
\vspace{.3em}
7&1.9&3.1&16.5&7.3$\cdot10^{11}$\\ 
\vspace{.3em}
19&2.5&2.3&14.2&8.5$\cdot10^{11}$ \\ 
\vspace{.3em}
34&2.2&2.7&15&8$\cdot10^{11}$ \\
\vspace{.3em}
49&1.8&3.3&10&1.2$\cdot10^{12}$\\
\vspace{.3em}
64&2.2&2.7&7&1.7$\cdot10^{12}$\\
\vspace{.3em}
72&2.3&2.6&8&1.5$\cdot10^{12}$
\end{tabular}
\caption{\label{table1} Summary of performance data in the TDIC using the Lenz lens resonator. The average pressure, $p$, was obtained at the centre of the 250 $\mu m$ culeted diamond anvil. The 90 degree pulse lengths, $t_{\pi/2}$, were obtained by nutation experiments at 1 W pulse power, and were used to estimate the average r.f. magnetic field strengths $\langle B_1\rangle$. LOD$_t$ was estimated given the obtained time domain signal-to-noise ratio, SNR$_t$, a number of approximately 1.7$\cdot10^{16}$ hydrogen nuclei in the sample cavity and a receiver bandwidth of 2 MHz at all measurements.}
\end{table}
Figure \ref{fig567}c) summarises recorded single scan $^1$H-NMR spectra of paraffin, and figure \ref{fig567}d) shows the evolution of the full width half maximum (FWHM) line widths. In the liquid phase, FWHM was found to increase from 5 ppm at ambient pressure ($10^{-4}$ GPa) to about 15 ppm at 7 GPa. At higher pressures, above the amorphisation pressure of paraffin oil\cite{Otto1998, Tateiwa2009, Tateiwa2010} at about 10 \-- 12 GPa, the line widths begin to increase exponentially reaching 23 ppm at 72 GPa. This effect is consistent with previous investigations of pressure gradients in the sample hole using paraffin oil as a pressure medium, i.e. see figure 3 from ref. [26]\cite{Meier2015a}  or figure 19 from ref. [22]\cite{Meier2017}. Also, as the typical chemical shift ranges of CH$_2$ and CH$_3$ groups are well below 5 ppm\cite{Boschke1983}, non-hydrostatic line broadening effects above the glass transition are within the observed line widths.\\
Nevertheless, such relatively narrow proton spectra indicate a non-vanishing mobility of the methylene and methyl groups in paraffin. Similar dynamic effects could be observed in molecular hydrogen in guest matricies under elevated pressures\cite{Okuchi2005a, Okuchi2007, Okuchi2011}. At the highest pressure, a doublet powder pattern developed, which can be associated with a gradual diminishing of molecular mobility and a resulting Pake doublet\cite{Pake1948}.
To investigate this effect further, spin-lattice (T$_1$) and spin-spin (T$_2$) relaxation measurements would illuminate the contribution of rotational and diffusional motion of the $^1$H nuclei, but as the main purpose of this study was to test the feasibility of the Lenz lenses, this effect has not been investigated further. \\
 Thus it can be assumed that the number of spins contributing to the observed sharp NMR signal stays constant, and limits of detection could be calculated using eq.\ref{LOD} and were found  to be almost constant, with an increase from $6\cdot10^{11} ~\text{spins}/{\sqrt{\text{Hz}}}$ at ambient conditions up to $1.5\cdot10^{12}~\text{spins}/{\sqrt{\text{Hz}}}$ at 72 GPa, see table \ref{table1}.\\
 \begin{figure}[htbp]
 \centering
 \includegraphics[scale=.40]{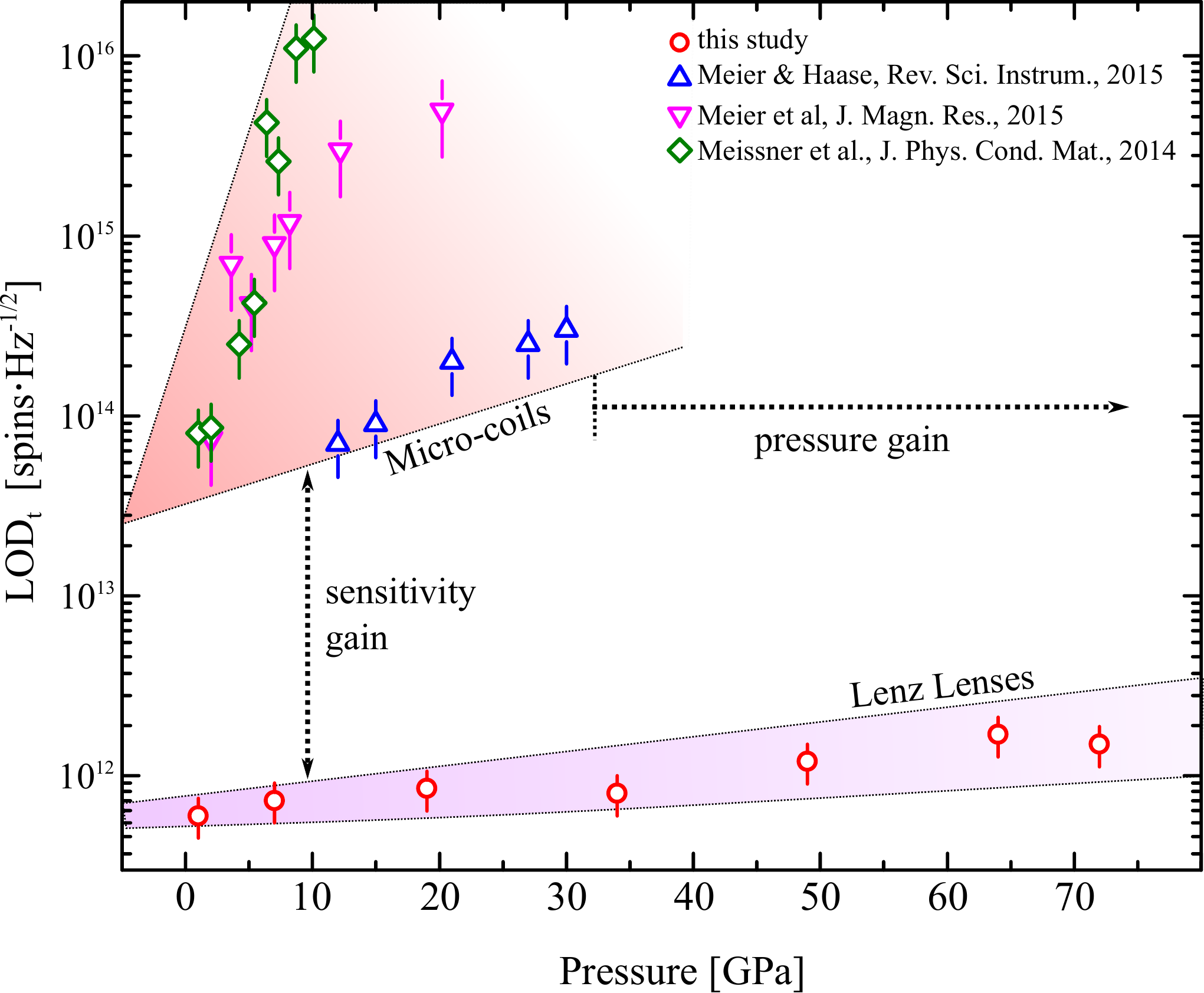}%
 \caption{Time domain limits of detection at increasing pressures obtained from micro-coil experiments and from Lenz lenses. The obtained gains in sensitivity and reachable pressures are indicated by black arrows. \label{fig8}}%
 \end{figure}
Figure \ref{fig8} compares the resulting LOD$_t$ of the Lenz lens resonator with three experimental runs using solenoidal micro-coils in the sample chamber, reaching maximal pressures of 10 GPa\cite{Meissner2014}, 20 GPa\cite{Meier2015} and 30 GPa\cite{Meier2015a}. 
In these experiments, the micro-coils range between 300 \-- 500 $\mu m$ in diameter and 100 \-- 250 $\mu m$ in height, tightly fitting into the initial sample volume. Evidently, the micro-coils' sensitivity exhibits a strong pressure dependence, with sensitivity losses as high as $10^{15}~\text{spins}/(\sqrt{\text{Hz}}\cdot \text{GPa})$ culminating in serious degradations of up to two orders of magnitude compared to their initial perfomances at ambient conditions. \\
Strikingly, detection limits of the Lenz lens set-up were not only found to be several orders of magnitude lower, and thus more sensitive, but also very stable, with small sensitivity losses of about $2\cdot 10^{10} ~\text{spins}/(\sqrt{\text{Hz}}\cdot \text{GPa})$ throughout the whole pressure run.\\
Such a significant difference between both resonator set-ups cannot be solely explained by the much smaller diameters of the resonators used, such an effect would only account for an increase of a factor of about five in the given experimetnal set-ups. Thus, in order to elucidate this problem further, numerical simulations have been performed, using the FEMM software package to simulate r.f. magnetic field maps for both resonator set-ups. Figure \ref{fig9} shows the results of the simulations comparing the initial magnetic fields prior to pressurisation with a significantly deformed arrangement.\\
\begin{figure*}[htbp]
  \includegraphics[scale=.35]{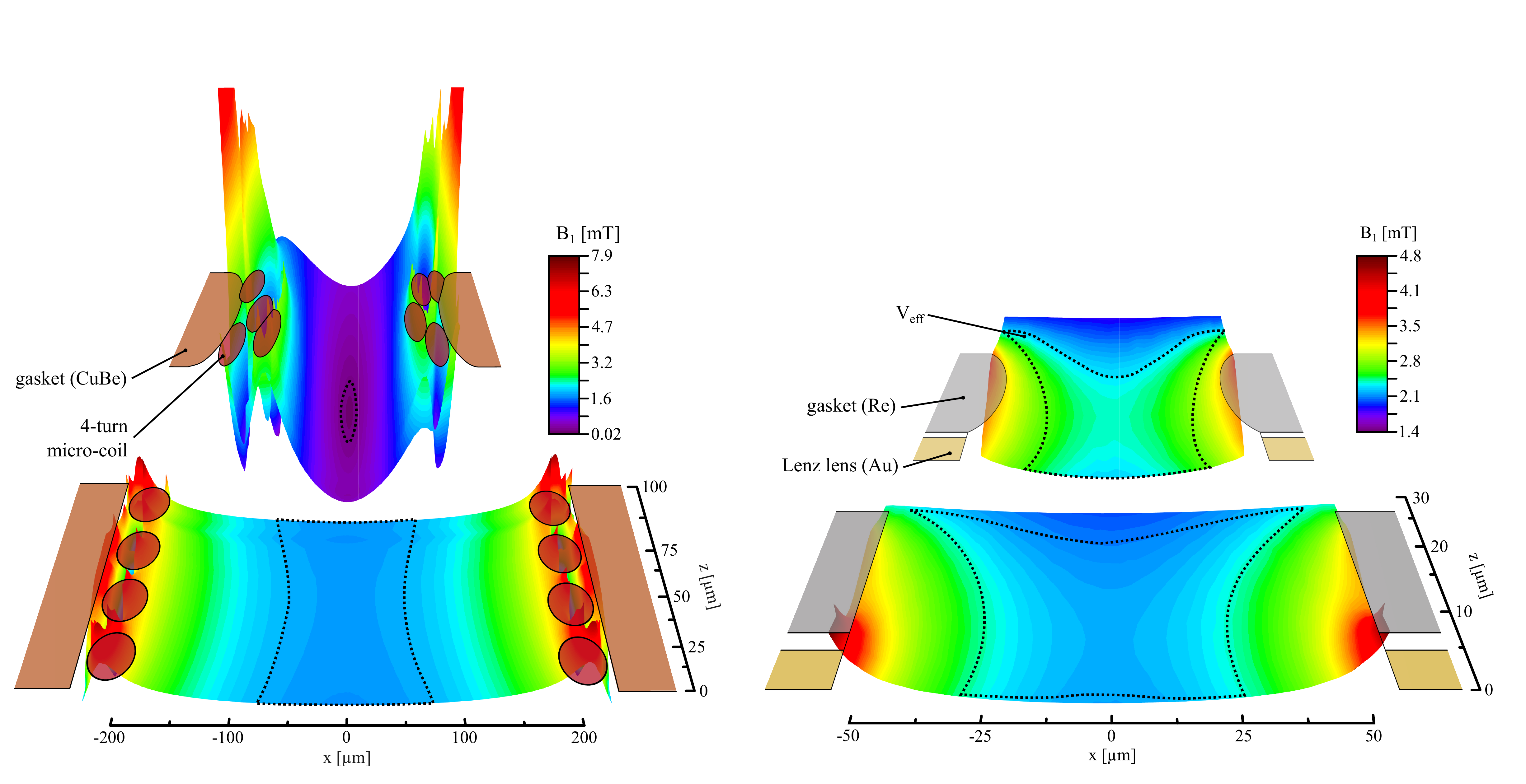}%
 \caption{ \label{fig9} Magnetic field maps of the B$_1$ fields generated by a micro-coil of four turns (400$\mu m$ in diameter and 100 $\mu m$ in height, left), and of a flat Lenz lens made from a solid sheet of gold foil (right). Above the initial configurations, the figure shows two deformed set-ups at a stage where the initial height and diameter of the sample cavity are reduced by 50\%, occuring well below 10 GPa when bigger CuBe gaskets were used\cite{Meier2016}. Using flat rhenium gaskets, this state of deformation typically occurs at substantially higher pressures. Indicated as an overlay are the geometries of both r.f. resonators and the corresponding gaskets. The deformed state of the micro-coil was reproduced from a photograph of an opened pressure cell working up to 6 GPa. Different gasket materials where used in both set-ups, which were also included in the simulations.  The initial parameters \--- operating frequency and circuit currents \--- where adjusted to the experimental parameters with which the spectra of figure \ref{fig567}c where recorded. The dotted lines represent the effective observable sample volume V$_\text{eff}$.}
 \end{figure*}
In accordance with similar calculations from van Bentum et al.\cite{VanBentum2007}, the B$_1$ field map in the x-z plane of a flat micro-coil, with a length-to-diameter ratio of less than unity, the magnetic field is rather inhomogeneously distributed with the highest magnetic fields close to the respective windings. The effective observable sample volume, V$_{eff}$, with a B$_1$ homogeneity within 20 \% of the central field, accounts to about 1.7 nl, which is about 14 \% of the total available sample space, and holds only 6\% of the total stored magnetic field energy of the coil.
 Moreover, as indicated by the \grqq deformed resonator\grqq, the B$_1$ field homogeneity greatly suffers from an irregular arrangement of the current carrying wire segments of the micro-coil. Such a deformed state typically arises already at relatively low pressures. Depending on the choice of gasket materials, the sample volume will be halved in its initial diameter and height within the range of some  few GPa. In this particular case, V$_{eff}$ drops to a twentieth of a percent due to significant B$_1$ field inhomogeneities, while at the same time storing only about 0.003 \% of the total magnetic field energy. Also, at such compressions, the risk for coil-gasket or inter-turn short circuits increases rapidly\cite{Meier2016}, rendering an application of micro-coils in DACs increasingly unreliable above 10  GPa. This effect becomes even more evident if the actual micro-coil geometries used in the corresponding experiments are considered. Meissner et al. used a micro-coil made from d=10 $\mu m$ insulated (plus 5 $\mu m$ insulation layer) copper wire, consisting of N=10 turns\cite{Meissner2012}. The total height of the coil can be gauged, using h$\approx$1.5Nd\cite{Peck1995}, to be 225 $\mu m$, which is already about 50 $\mu m$ higher than the initial sample chamber drilled into the CuBe gasket used. For the study reaching 20 GPa, Meier et al. used micro-coils consisting of 4 to 5 turns using insulated copper wire of 23 $\mu m$ thickness (140 $\mu m$ to 170 $\mu m$ total height ). As can be seen the pressure induced degradation is reduced by about half an order of magnitude compared to the pressure run reaching only 10 GPa. Finally, in the experiments reaching 30 GPa, micro-coils of only 70 $\mu m$ could be employed with considerably reduced sample chambers. Evidently, detection limits could be improved considerably by at least one order of magnitude, but still did not permit for very high sensitivity NMR above this pressure.\\
 In the case of the Lenz lenses, on the other hand, the r.f. B$_1$ field appears to be rather homogeneous over at least 40\--50 \% of the total sample cavity, storing about 30 \% of the magnetic field energy. Strikingly, under compression, the situation does not deteriorate significantly, and both the stored energy ($\approx 35 \%$) and V$_{\text{eff}}$ ($\approx 47 \%$) remain almost constant. The B$_1$ field strengths of the lens resonators found in the simulations compare well with the actual field strengths found via nutation experiments (see table \ref{table1}), which is further evidence of the applicability of this approach.\\
 
 \section{Discussion}
This study demonstrates that nuclear magnetic resonance spectroscopy can be implemented in modern high pressure vessels at formerly unprecedented pressures using a new type of r.f. resonator. The advantages of this approach compared to formerly used high pressure resonators are manifold.\\
Lenz lenses allow for a significantly higher spin sensitivity, and thus excellent limits of detection in the order of $10^{12}~\text{spins}/\sqrt{\text{Hz}}$, which corresponds to a detectable volume of only 15 $\mu m^3$ of a water sample. Such high sensitivities are ideal for applications suffering from highly limited sample volumes that barely fit in the limited space of a diamond anvil cell.  Moreover, as the sensitivity does not deteriorate with pressure, very high sensitivity NMR at 1 Mbar and beyond becomes possible, even more so as the r.f. B$_1$ field and hence the mass sensitivity can be further increased when the central hole of the lens is reduced.\\
 Secondly, and perhaps more importantly, the Lenz lenses constitute the first quasi two-dimensional r.f. resonators used for high pressure applications, resulting in substantially higher pressures as the overall pressure cell set-up is not altered significantly, compared to micro-coil set-ups where small grooves in the metallic gasket are mandatory to avoid premature coil-gasket short cuts.\\
Finally, this study also demonstrates that $^1$H-NMR is possible without the interference of spurious signals \---typically originating from organic materials close to the resonator such as wire insulations or epoxy resins\--- which are prone to distort or completely mask the weak proton signal originating from within the sample chamber. Thus, it now becomes possible to study
materials at formerly unprecedented energy densities, potentially yielding surprising new phenoma in high pressure chemistry and biophysics. \\
The paramount significance for bio-chemistry and life sciences becomes obvious if the current state-of-the-art high pressure NMR technique used for investigating proteins\cite{Ballard1997, Kremer2006} is considered. This technique uses rather large r.f. resonators in an autoclave system, with sample volumes in the order of 10 \-- 100 $\mu l$, yielding sensitivities of about $10^{18}\--10^{19}~\text{spins}/\sqrt{\text{Hz}}$ at considerably lower maximal pressures of less than 1 GPa\cite{Ballard1997, Arnold2003}. Therefore, the application of Lenz lenses in modern high pressure vessels could lead to a renewed interest of NMR investigations of protein folding dynamics at considerably higher pressures than possible before.

\begin{acknowledgments}
We would like to thank Professor Roessler for provision of the 9.3 T NMR system. Furthermore, we thank Nobuyoshi Miyajima and Katharina Marquardt for their help with the FIB and the ion
milling (grant number: INST 90/315-1 FUGG). We are also very thankful for the help of Sven Linhardt and Stefan Uebelhack for manufactoring the NMR probe and pressure cell components. Klaus Mueller performed the sputter coating.
 The authors, T.M., S.P. and L.D., were funded by the Bavarian Geoinstitute through the Free State of Bavaria. J.G.K. was supported by the Senior Grant 290586 of the European Research Council. N.W. and D.M. were funded through the German Research Foundation, DFG-RUMS project (KO-1883-23-1).
\end{acknowledgments}

\section{Author Contributions}

T.M., L.D. and S.P. prepared the pressure cells. S.P. conducted the ion milling of the anvils. T.M., N.W., D.M. and J.G.K. designed the resonators. T.M. devised and conducted the experiments. T.M. performed the data analysis and magnetic field simulations. T.M. and L.D. wrote the manuscript.

\bibliographystyle{unsrtnat} 

\begin{thebibliography}{57}
\providecommand{\natexlab}[1]{#1}
\providecommand{\url}[1]{\texttt{#1}}
\expandafter\ifx\csname urlstyle\endcsname\relax
  \providecommand{\doi}[1]{doi: #1}\else
  \providecommand{\doi}{doi: \begingroup \urlstyle{rm}\Url}\fi

\bibitem[Grant and Harris(2007)]{Grant2007}
D.~M. Grant and R.~K. Harris.
\newblock \emph{{Encyclopedia of Nuclear Magnetic Resonance}}.
\newblock John Wiley {\&} Sons, Ltd, Chichester, UK, 2007.
\newblock ISBN 978-0-471-49082-1.

\bibitem[Ernst(2010)]{Ernst2010}
R.~R. Ernst.
\newblock \emph{Angewandte Chemie International Edition}, 49\penalty0
  (45):\penalty0 8310--8315, nov 2010.
\newblock \doi{10.1002/anie.201005067}.

\bibitem[Andrew(1984)]{Andrew1984}
E.~R. Andrew.
\newblock \emph{British Medical Bulletin}, 40\penalty0 (2):\penalty0 115--119,
  1984.

\bibitem[Lauterbur(1973)]{Lauterbur1973}
P.~C. Lauterbur.
\newblock \emph{Nature (London, United Kingdom)}, 242:\penalty0 190--191, 1973.
\newblock \doi{10.1038/242190a0}.

\bibitem[Mansfield(2004)]{Mansfield2004}
P. Mansfield.
\newblock \emph{Angewandte Chemie - International Edition}, 43\penalty0
  (41):\penalty0 5456--5464, 2004.
\newblock \doi{10.1002/anie.200460078}.

\bibitem[Riek et~al.(1996)Riek, Hornemann, Wider, Billeter, Glockshuber, and
  W{\"{u}}thrich]{Riek1996}
R.~Riek, S.~Hornemann, G.~Wider, M.~Billeter, R.~Glockshuber, and
  K.~W{\"{u}}thrich.
\newblock \emph{Nature}, 382\penalty0 (6587):\penalty0 180--182, jul 1996.
\newblock \doi{10.1038/382180a0}.

\bibitem[Zahn et~al.(2000)Zahn, Liu, L{\"{u}}hrs, Riek, von Schroetter,
  {L{\'{o}}pez Garc{\'{i}}a}, Billeter, Calzolai, Wider, and
  W{\"{u}}thrich]{Zahn2000}
R.~Zahn, A.~Liu, T.~L{\"{u}}hrs, R.~Riek, C.~von Schroetter, F.~{L{\'{o}}pez
  Garc{\'{i}}a}, M.~Billeter, L.~Calzolai, G.~Wider, and K.~W{\"{u}}thrich.
\newblock \emph{Proceedings of the National Academy of Sciences of the United
  States of America}, 97\penalty0 (1):\penalty0 145--50, 2000.
\newblock \doi{10.1073/pnas.97.1.145}.

\bibitem[Scheidt et~al.(2011)Scheidt, Sickert, Meier, Castellucci, Tomasini,
  and Huster]{Scheidt2011}
H. Scheidt, A. Sickert, T. Meier, N. Castellucci, C.Tomasini, and D. Huster.
\newblock \emph{Organic {\&} biomolecular chemistry}, 9\penalty0 (20):\penalty0
  6998--7006, 2011.
\newblock \doi{10.1039/c1ob05652b}.

\bibitem[Yarger et~al.(1995)Yarger, Smith, Nieman, Diefenbacher, Wolf, Poe, and
  McMillan]{Yarger1995a}
J.~L. Yarger, K.~H. Smith, R.~A. Nieman, J.~Diefenbacher, G.~H. Wolf, B.~T.
  Poe, and P.~F. McMillan.
\newblock \emph{Science}, 270\penalty0 (5244):\penalty0 1964--1967, dec 1995.
\newblock \doi{10.1126/science.270.5244.1964}.

\bibitem[Stebbins(1995)]{Stebbins1995}
J.~F. Stebbins.
\newblock In \emph{Structure, Dynamics and Properties of Silicate Melts}, pages
  191--246. Washington D.C., 1 edition, 1995.

\bibitem[Sharma et~al.(2002)Sharma, Scott, Cody, Fogel, Hazen, Hemley, and
  Huntress]{Sharma2002}
A.~Sharma, J.~H. Scott, G.~D. Cody, M.~L. Fogel, R.~M. Hazen, R.~J. Hemley, and
  W.~T. Huntress.
\newblock \emph{Science}, 295\penalty0 (5559):\penalty0 1514--1516, feb 2002.
\newblock \doi{10.1126/science.1068018}.

\bibitem[Daniel et~al.(2006)Daniel, Oger, and Winter]{Daniel2006}
I.~Daniel, P.~Oger, and R.~Winter.
\newblock \emph{Chem Soc Rev}, 35\penalty0 (10):\penalty0 858--875, 2006.
\newblock \doi{10.1039/b517766a}.

\bibitem[Huang et~al.(2016)Huang, Tran, Rodgers, Bartlett, Hemley, and
  Ichiye]{Huang2016}
Q.~Huang, K.~N. Tran, J.~M. Rodgers, D.~H. Bartlett, R.~J. Hemley, and
  T.~Ichiye.
\newblock \emph{Condensed Matter Physics}, 19\penalty0 (2):\penalty0 1--16,
  2016.
\newblock ISSN 22249079.
\newblock \doi{10.5488/CMP.19.22801}.

\bibitem[Bridgman(1914)]{Bridgman1914}
P.~W. Bridgman.
\newblock \emph{Journal of Biological Chemistry}, 19\penalty0 (4):\penalty0
  511--512, 1914.

\bibitem[Balny(2006)]{Balny2006}
C.~Balny.
\newblock \emph{Biochimica et Biophysica Acta - Proteins and Proteomics},
  1764\penalty0 (3):\penalty0 632--639, 2006.
\newblock ISSN 15709639.
\newblock \doi{10.1016/j.bbapap.2005.10.004}.

\bibitem[Heremans and Smeller(1998)]{Heremans1998}
K.~Heremans and L.~Smeller.
\newblock \emph{Biochimica et biophysica acta}, 1386\penalty0 (2):\penalty0
  353--370, 1998.
\newblock \doi{10.1016/S0167-4838(98)00102-2}.

\bibitem[Silva et~al.(2001)Silva, Foguel, and Royer]{Silva2001}
J.~L.~Silva, D.~Foguel, and C.~A.~Royer.
\newblock \emph{Trends in Biochemical Sciences}, 26\penalty0 (10):\penalty0
  612--618, 2001.
\newblock \doi{10.1016/S0968-0004(01)01949-1}.

\bibitem[Mozhaev et~al.(1996)Mozhaev, Heremans, Frank, Masson, and
  Balny]{Mozhaev1996}
V.V.~Mozhaev, K.~Heremans, J.~Frank, P.~Masson, and C.~Balny.
\newblock \emph{Proteins}, 24\penalty0 (1):\penalty0 81--91, 1996.
\newblock \doi{10.1002/(SICI)1097-0134(199601)24:1<81::AID-PROT6>3.0.CO;2-R}.

\bibitem[Royer(2002)]{Royer2002}
C.A.~Royer.
\newblock \emph{Biochimica et Biophysica Acta - Protein Structure and Molecular
  Enzymology}, 1595\penalty0 (1-2):\penalty0 201--209, 2002.
\newblock \doi{10.1016/S0167-4838(01)00344-2}.

\bibitem[Golub et~al.(2017)Golub, Lehofer, Martinez, Ollivier, Kohlbrecher,
  Prassl, and Peters]{Golub2017}
M.~Golub, B.~Lehofer, N.~Martinez, J.~Ollivier, J.~Kohlbrecher, R.~Prassl, and
  J.~Peters.
\newblock \emph{Scientific Reports}, 7\penalty0 (April):\penalty0 46034, 2017.
\newblock \doi{10.1038/srep46034}.

\bibitem[Fukushima and Roeder(1994)]{Fukushima1994}
E.~Fukushima and S.B.W.~Roeder.
\newblock \emph{{Experimental Pulse NMR - A Nuts and Bolts Approach}}.
\newblock Addison Wesley, Reading, first edition, 1994.

\bibitem[Meier(2017)]{Meier2017}
T.~Meier.
\newblock \emph{Annual Reports on NMR Spectroscopy}, 94\penalty0, 2017., t.b.p.

\bibitem[Suzuki et~al.(2009)Suzuki, Yamauchi, Shimizu, Itoh, Takeshita,
  Terakura, Takagi, Tokura, Yamauchi, and Ueda]{Suzuki2009}
T.~Suzuki, I.~Yamauchi, Y.~Shimizu, M.~Itoh, N.~Takeshita, C.~Terakura,
  H.~Takagi, Y.~Tokura, T.~Yamauchi, and Y.~Ueda.
\newblock \emph{Physical Review B}, 79[1] T. S\penalty0 (8):\penalty0 081101,
 2009.
\newblock \doi{10.1103/PhysRevB.79.081101}.

\bibitem[Meier et~al.(2014)Meier, Herzig, and Haase]{Meier2014}
T.~Meier, T.~Herzig, and J.~Haase.
\newblock \emph{The Review of scientific instruments}, 85\penalty0
  (4):\penalty0 043903, 2014.
\newblock \doi{10.1063/1.4870798}.

\bibitem[Meier et~al.(2015)Meier, Reichardt, and Haase]{Meier2015}
T.~Meier, S.~Reichardt, and J.~Haase.
\newblock \emph{Journal of Magnetic Resonance}, 257:\penalty0 39--44, 2015.
\newblock \doi{10.1016/j.jmr.2015.05.007}.

\bibitem[Meier and Haase(2015)]{Meier2015a}
T.~Meier and J.~Haase.
\newblock \emph{Review of Scientific Instruments}, 86\penalty0 (12):\penalty0
  123906, 2015.
\newblock \doi{10.1063/1.4939057}.

\bibitem[Meier and Haase(2014)]{Meier2014a}
T.~Meier and J.~Haase.
\newblock \emph{Journal of Visualized Experiments}, \penalty0 (92):\penalty0
  1--10, 2014.
\newblock \doi{10.3791/52243}.

\bibitem[Schwarz et~al.(1999)Schwarz, Grzechnik, Syassen, Loa, and
  Hanfland]{Schwarz1999}
U.~Schwarz, A.~Grzechnik, K.~Syassen, I.~Loa, and M.~Hanfland.
\newblock \emph{Physical Review Letters}, 83\penalty0 (20):\penalty0
  4085--4088, 1999.
\newblock \doi{10.1103/PhysRevLett.83.4085}.

\bibitem[Bykova et~al.(2016)Bykova, Dubrovinsky, Dubrovinskaia, Bykov,
  McCammon, Ovsyannikov, Liermann, Kupenko, Chumakov, R{\"{u}}ffer, Hanfland,
  and Prakapenka]{Bykova2016}
E.~Bykova, L.~Dubrovinsky, N.~Dubrovinskaia, M.~Bykov, C.~McCammon, S.V.~Ovsyannikov, H.P.~Liermann, I.~Kupenko, A.I.~Chumakov, R.~R{\"{u}}ffer,
  M.~Hanfland, and V.~Prakapenka.
\newblock \emph{Nature Communications}, 7:\penalty0 10661, 2016.
\newblock \doi{10.1038/ncomms10661}.

\bibitem[Frost et~al.(2004)Frost, Liebske, Langenhorst, McCammon, Tronnes, and
  Rubie]{Frost2004}
D.J.~Frost, C.~Liebske, F.~Langenhorst, C.~McCammon, R.~G. Tronnes, and D.C.~Rubie.
\newblock \emph{Nature}, 428:\penalty0 409--412, 2004.
\newblock \doi{10.1029/200/JC000964}.

\bibitem[Kupenko et~al.(2015)Kupenko, McCammon, Sinmyo, Cerantola, Potapkin,
  Chumakov, Kantor, R{\"{u}}ffer, and Dubrovinsky]{Kupenko2015a}
I.~Kupenko, C.~McCammon, R.~Sinmyo, V.~Cerantola, V.~Potapkin, A.I.~Chumakov,
  A.~Kantor, R.~R{\"{u}}ffer, and L.~Dubrovinsky.
\newblock \emph{Earth and Planetary Science Letters}, 423:\penalty0 78--86,
  2015.
\newblock \doi{10.1016/j.epsl.2015.04.027}.

\bibitem[Kurnosov et~al.(2017)Kurnosov, Marquardt, Frost, Ballaran, and
  Ziberna]{Kurnosov2017}
A.~Kurnosov, H.~Marquardt, D.J.~Frost, T.~Boffa Ballaran, and L.~Ziberna.
\newblock \emph{Nature}, 543\penalty0 (7646):\penalty0 543--546, 2017.
\newblock \doi{10.1038/nature21390}.

\bibitem[Eremets and Trojan(2011)]{Eremets2011}
M.I.~Eremets and I.A.~Trojan.
\newblock \emph{Nature Materials}, 10\penalty0 (12):\penalty0 927--931,
  2011.
\newblock \doi{10.1038/nmat3175}.

\bibitem[Eremets et~al.(2016)Eremets, Troyan, and Drozdov]{Eremets2016}
M.I.~Eremets, I.A.~Troyan, and A.P.~Drozdov.
\newblock {arXiv:1601.04479}.
\newblock  2016.

\bibitem[Dalladay-Simpson et~al.(2016)Dalladay-Simpson, Howie, and
  Gregoryanz]{Dalladay-Simpson2016}
P.~Dalladay-Simpson, R.T.~Howie, and E.~Gregoryanz.
\newblock \emph{Nature}, 529\penalty0 (7584):\penalty0 63--67, jan 2016.
\newblock \doi{10.1038/nature16164}.

\bibitem[Dias and Silvera(2017)]{Dias2017}
R.P.~Dias and I.F.~Silvera.
\newblock \emph{Science}, page 1579, 2017.
\newblock \doi{10.1126/science.aal1579}.

\bibitem[Schoenmaker et~al.(2013)Schoenmaker, Pirota, and
  Teixeira]{Schoenmaker2013}
J.~Schoenmaker, K.R.~Pirota, and J.C.~Teixeira.
\newblock \emph{Review of Scientific Instruments}, 84\penalty0 (8):\penalty0
  085120, 2013.
\newblock \doi{10.1063/1.4819234}.

\bibitem[Spengler et~al.(2016)Spengler, While, Meissner, Wallrabe, and
  Korvink]{Spengler2016}
N.~Spengler, P.T.~While, M.V.~Meissner, U.~Wallrabe, and J.G.~Korvink.
\newblock {arXiv:1606.07044}.
\newblock pages 1--18, 2016.

\bibitem[Ryan et~al.(2012)Ryan, Song, Za{\ss}, Korvink, and Utz]{Ryan2012}
H.~Ryan, S.H.~Song, A.~Za{\ss}, J.~Korvink, and M.~Utz.
\newblock \emph{Analytical Chemistry}, 84\penalty0 (8):\penalty0 3696--3702,
  2012.
\newblock \doi{10.1021/ac300204z}.

\bibitem[Pravica and Silvera(1998)]{Pravica1998}
M.G.~Pravica and I.F.~Silvera.
\newblock \emph{Review of Scientific Instruments}, 69\penalty0 (2):\penalty0
  479--484, 1998.
\newblock \doi{10.1063/1.1148686}.

\bibitem[Okuchi et~al.(2005{\natexlab{a}})Okuchi, Hemley, and Mao]{Okuchi2005}
T.~Okuchi, R.J.~Hemley, and H.-k.~Mao.
\newblock \emph{Review of Scientific Instruments}, 76\penalty0 (2):\penalty0
  026111, 2005{\natexlab{a}}.
\newblock \doi{10.1063/1.1850653}.

\bibitem[Otto et~al.(1998)Otto, Vassiliou, and Frommeyer]{Otto1998}
J.W.~Otto, J.K.~Vassiliou, and G.~Frommeyer.
\newblock \emph{Physical Review B}, 57\penalty0 (6):\penalty0 3253--3263, 1998.
\newblock \doi{10.1103/PhysRevB.57.3253}.

\bibitem[Tateiwa and Haga(2009)]{Tateiwa2009}
N.~Tateiwa and Y.~Haga.
\newblock \emph{Review of Scientific Instruments}, 80\penalty0 (12):\penalty0
  123901/1----/7, 2009.
\newblock \doi{10.1063/1.3265992}.

\bibitem[Tateiwa and Haga(2010)]{Tateiwa2010}
N.~Tateiwa and Y.~Haga.
\newblock \emph{Journal of Physics: Conference Series}, 215:\penalty0 12178,
  2010.
\newblock \doi{10.1088/1742-6596/215/1/012178}.

\bibitem[Boschke et~al.(1983)Boschke, Fresenius, Huber, Pungor, Rechnitz,
  Simon, and West]{Boschke1983}
F.L.~Boschke, W.~Fresenius, J.F.K.~Huber, E.~Pungor, G.A.~Rechnitz,
  W.~Simon, and T.S.~West.
\newblock \emph{{Spectral Data for Structure Determination of Organic
  Compounds}}.
\newblock Springer, Berlin, Heidelberg, first edition, 1983.

\bibitem[Okuchi et~al.(2005{\natexlab{b}})Okuchi, Cody, Mao, and
  Hemley]{Okuchi2005a}
T.~Okuchi, G.D.~Cody, H.-k.~Mao, and R.J.~Hemley.
\newblock \emph{Journal of Chemical Physics}, 122\penalty0 (24):\penalty0
  244509, 2005{\natexlab{b}}.
\newblock \doi{10.1063/1.1944732}.

\bibitem[Okuchi et~al.(2007)Okuchi, Takigawa, Shu, Mao, Hemley, and
  Yagi]{Okuchi2007}
T.~Okuchi, M.~Takigawa, J.~Shu, H.-k.~Mao, R.J.~Hemley, and T.~Yagi.
\newblock \emph{Physical Review B - Condensed Matter and Materials Physics},
  75\penalty0 (14):\penalty0 144104, 2007.
\newblock \doi{10.1103/PhysRevB.75.144104}.

\bibitem[Okuchi(2011)]{Okuchi2011}
T.~Okuchi.
\newblock \emph{The Journal of Physical Chemistry C}, 116\penalty0
  (3):\penalty0 2179--2182, 2011.
\newblock \doi{10.1021/jp206732f}.

\bibitem[Pake(1948)]{Pake1948}
G.E.~Pake.
\newblock \emph{The Journal of Chemical Physics}, 16\penalty0 (4):\penalty0
  327--336, 1948.
\newblock \doi{10.1063/1.1746878}.

\bibitem[Meissner et~al.(2014)Meissner, Goh, Haase, Richter, Koepernik, and
  Eschrig]{Meissner2014}
T.~Meissner, S.K.~Goh, J.~Haase, M.~Richter, K.~Koepernik, and H.~Eschrig.
\newblock \emph{Journal of Physics: Condensed Matter}, 26\penalty0
  (1):\penalty0 015501, 2014.
\newblock \doi{10.1088/0953-8984/26/1/015501}.

\bibitem[Meier(2016)]{Meier2016}
T.~Meier.
\newblock \emph{{High Sensitivity Nuclear Magnetic Resonance at Extreme
  Pressures}}.
\newblock PhD thesis, Leipzig University, 2016.

\bibitem[van Bentum et~al.(2007)van Bentum, Janssen, Kentgens, Bart, and
  Gardeniers]{VanBentum2007}
P.J.M.~van Bentum, J.W.G.~Janssen, A.P.M.~Kentgens, J.~Bart, and J.G.E.~Gardeniers.
\newblock \emph{Journal of Magnetic Resonance}, 189\penalty0 (1):\penalty0
  104--113, 2007.
\newblock \doi{10.1016/j.jmr.2007.08.019}.

\bibitem[Meissner(2013)]{Meissner2012}
T.~Meissner.
\newblock \emph{{Exploring Nuclear Magnetic Resonance at the Highest
  Pressures}}.
\newblock PhD thesis, Leipzig University, 2013.

\bibitem[Peck et~al.(1995)Peck, Magin, and Lauterbur]{Peck1995}
T.L.~Peck, R.L.~Magin, and P.C.~Lauterbur.
\newblock \emph{Journal of magnetic resonance. Series B}, 108\penalty0
  (2):\penalty0 114--124, 1995.
\newblock \doi{10.1006/jmrb.1995.1112}.

\bibitem[Ballard and Jonas(1997)]{Ballard1997}
L.~Ballard and J.~Jonas.
\newblock \emph{Annual Reports on NMR Spectroscopy}, 33:\penalty0 115--150,
  1997.

\bibitem[Kremer(2006)]{Kremer2006}
W.~Kremer.
\newblock \emph{Annual Reports on NMR Spectroscopy}, 57:\penalty0 177--200,
  2006.

\bibitem[Arnold et~al.(2003)Arnold, Kalbitzer, and Kremer]{Arnold2003}
M.R.~Arnold, H.R.~Kalbitzer, and W.~Kremer.
\newblock \emph{Journal of Magnetic Resonance}, 161\penalty0 (2):\penalty0
  127--131, 2003.
\newblock \doi{10.1016/S1090-7807(02)00179-9}.

\end{thebibliography}

\end{document}